\documentclass[aps,prc,superscriptaddress,twoside,twocolumn,nofootinbib,10pt,%
showpacs,floatfix]{revtex4-1}

\usepackage{amsmath,amssymb}
\usepackage{graphicx,bm}
\usepackage{epstopdf}
\usepackage{ulem} 
\usepackage[usenames]{color}
\usepackage{float}

\allowdisplaybreaks

\newcommand{\Tr}{\textrm{Tr}}
\newcommand{\vsla}{v\hspace{-.47em}/}
\newcommand{\fpi}{f_{\pi}}
\newcommand{\non}{\nonumber}
\newcommand{\del}{\partial}

\begin{document}

\title{Heavy quark spin multiplet structure of $\bar{P}^{(*)}\Sigma_Q^{(*)}$ molecular states}

\author{Yuki Shimizu}
\email{yshimizu@hken.phys.nagoya-u.ac.jp}
\affiliation{Department of Physics,  Nagoya University, Nagoya 464-8602, Japan}

\author{Yasuhiro Yamaguchi}
\email{yasuhiro.yamaguchi@riken.jp}
\affiliation{Theoretical Research Division, Nishina Center, RIKEN, Hirosawa, Wako, Saitama 351-0198, Japan}

\author{Masayasu Harada}
\email{harada@hken.phys.nagoya-u.ac.jp}
\affiliation{Department of Physics,  Nagoya University, Nagoya 464-8602, Japan}

\date{\today}


\begin{abstract}
We study the structure of  heavy quark spin (HQS) multiplets for 
heavy meson-baryon molecular states in a coupled system of $\bar{P}^{(*)}\Sigma_{Q}^{(*)}$, 
with constructing the one-pion exchange potential with S-wave orbital angular 
momentum.  
Using the light cloud spin basis, 
we find that there are four types of HQS multiplets 
classified by the structure of heavy quark spin and light cloud spin. 
The multiplets which have attractive potential are determined 
by the sign of the coupling constant for the heavy meson-pion interactions. 
Furthermore, the difference in the structure of light cloud spin gives the restrictions of the decay channel, 
which implies that the partial decay width has the information for the structure of HQS multiplets.
This behavior is more likely to appear in hidden-bottom sector than in hidden-charm sector.
\end{abstract}

\maketitle


\section{Introduction}
\label{sec:Intro}
The exotic hadrons are the very interesting research subjects in hadron and nuclear physics.
In 2015, the Large Hadron Collider beauty experiment (LHCb) collaboration announced 
the observation of two hidden-charm pentaquarks, $P_c^+(4380)$ and $P_c^+(4450)$, 
in the decay of $\Lambda_b^0 \to J/\psi K^- p$ \cite{Aaij:2015tga, Aaij:2016phn, Aaij:2016ymb}.
Their masses are $M_{4380}=4380\pm8\pm28$ MeV and $M_{4450}=4449.8\pm1.7\pm2.5$ MeV, 
and decay widths are $\Gamma_{4380}=205\pm18\pm86$ MeV and $\Gamma_{4450}=39\pm5\pm19$ MeV.
The spin and parity $J^P$ of them are not well determined.
The one state is $J=3/2$ and the other state is $J=5/2$ and they have opposite parity.

Before the LHCb observation, some theoretical studies of hidden-charm pentaquarks were done~\cite{Wu:2010jy, Yang:2011wz, Wang:2011rga, Wu:2012md}.
After the LHCb announcement, there are many analyses based on the hadronic molecular state 
\cite{Chen:2015loa, He:2015cea, Chen:2015moa, Huang:2015uda, Roca:2015dva, Meissner:2015mza, Xiao:2015fia, Burns:2015dwa, Kahana:2015tkb, Chen:2016heh, Chen:2016otp, Shimizu:2016rrd, Yamaguchi:2016ote, He:2016pfa, Ortega:2016syt, Azizi:2016dhy, Geng:2017hxc}, 
compact pentaquark state
\cite{Maiani:2015vwa, Lebed:2015tna, Anisovich:2015cia, Li:2015gta, Wang:2015epa, Zhu:2015bba, Santopinto:2016pkp, Wu:2017weo, Hiyama:2018ukv},
quark-cluster model \cite{Takeuchi:2016ejt},
baryocharmonium model \cite{Kubarovsky:2015aaa},
hadroquarkonia model \cite{Eides:2017xnt},
topologial soliton model \cite{Scoccola:2015nia},
and meson-baryon molecules coupled with five-quark states \cite{Yamaguchi:2017zmn}.
The kinematical rescattering effects are also
discussed in Refs.~\cite{Guo:2015umn, Liu:2015fea, Mikhasenko:2015vca, Guo:2016bkl, Bayar:2016ftu}.

There are many theoretical descriptions for $P_c^+$ pentaquarks.
Among those pictures, the hadronic molecular one
has been used for several other exotic hadrons, especially near the thresholds.
For example, since the mass of $X(3872)$ is close to the $D\bar{D}^\ast$ threshold, $X(3872)$ includes the $D\bar{D}^*$ molecule structure~\cite{Takeuchi:2014rsa}.
The masses of $P_c^+(4380)$ and $P_c^+(4450)$ are
slightly below the thresholds of $\bar{D}\Sigma_c^*$ and $\bar{D}^*\Sigma_c$, respectively.
They can be considered as the loosely bound state of heavy meson and heavy baryon.

Charm quarks are included in $P_c^+$ pentaquarks.
The masses of the heavy quarks, charm and bottom, are much larger than the typical scale
of low energy QCD, $\Lambda_{\rm QCD} \sim 200$MeV.
For the heavy quark region,
there is a characteristic property in the quark interaction.
The spin dependent interaction of the heavy quark is suppressed by the inverse of the heavy quark mass, $1/m_Q$.
By this suppression, heavy quark spin symmetry (HQS) is appeared in the heavy quark limit~\cite{Isgur:1989vq,Isgur:1989ed,Isgur:1991wq,Neubert:1993mb,Manohar:2000dt}.
As a result, we can decompose the total spin $\vec{J}$ to heavy quark spin $\vec{S}$ and the other spin $\vec{j}$~:
\begin{align}
	\vec{J} = \vec{S} + \vec{j}~.
\end{align}
The total spin is conserved and heavy quark spin is also conserved in the
heavy quark limit
because of the suppression of the spin dependent force.
Thus the other spin part is also conserved.
This conservation leads to the mass degeneracy of heavy hadrons.
Let us consider the heavy meson $q\bar{Q}$ with a light quark $q$ and a heavy quark $Q$.
For $j \geq 1/2$, there are two degenerate states with total spin
\begin{align}
	J_{\pm} = j \pm 1/2~.
\end{align}
These two states are called HQS doulblet.
There is only $J=1/2$ state for $j=0$, hence it is called HQS singlet.

Such HQS multiplet structure is seen in the charm and bottom hadron
mass spectrum.
For example, the small mass difference is obtained between the
heavy-light pseudoscalar ($J=0$) and vector ($J=1$) mesons, 140 MeV between $D$ and
$D^\ast$,
and 45 MeV between $B$ and $B^\ast$.
These mass splittings are much smaller than those in the light quark
sectors, 600 MeV between $\pi$ and $\rho$, and 400 MeV between $K$ and
$K^\ast$.
This observation indicates that the approximate heavy quark
spin symmetry is realized in the charm and bottom quark sectors, and these two mesons
with $J=0,1$ belong to the HQS doublet having the heavy spin $S=1/2$ and the other spin $j=1/2$.

The approximate mass degeneracy is also observed in the heavy-light baryons.
The mass splitting between $\Sigma_c$ ($J=1/2$) and $\Sigma^\ast_{c}$
($J=3/2$) ($\Sigma_b$ and $\Sigma^\ast_b$) is about 65 MeV (20 MeV).
They are the HQS doublet state with the heavy spin $S=1/2$ and the
other spin $j=1$.
On the other hand, the heavy-light baryons $\Lambda_c$ and
$\Lambda_b$ with the light diquark spin 0 are a HQS singlet state.

In this paper, we study the structure of HQS 
multiplets of $Q\bar{Q}qqq$-type pentaquarks 
regarding them as molecular states of $\bar{P}^{(*)}\Sigma_Q^{(*)}$.
Here, $\bar{P}^{(*)}$ means a HQS doublet meson with an anti-heavy quark like $\bar{D}(B)$ and $\bar{D}^*(B^*)$ and $\Sigma_Q^{(*)}$ stands for a HQS doublet baryon with a heavy quark like $\Sigma_c(\Sigma_b)$ and $\Sigma_c^*(\Sigma_b^*)$~.

The HQS doublet structures of $\bar{P}^{(*)}$ meson and $\Sigma_Q^{(*)}$ baryon which have one heavy quark are well known.
HQS multiplet structure of $\bar{P}^{(*)}N$ molecular state with a single heavy quark is discussed in Refs.~\cite{Yasui:2013vca, Yamaguchi:2014era,Hosaka:2016ypm}.
They showed that the degeneracy of $j \pm 1/2$ states can be expanded to multi-hadron system.
In this paper, we study the HQS multiplet structure of $P_c$-like pentaquarks as a doubly heavy quarks system.
The appearance of the HQS multiplet for
$\bar{P}^{(*)}\Sigma_Q^{(*)}$ molecules is demonstrated by introducing
the one pion exchange potential (OPEP) which is derived from the heavy 
hadron effective theory
respecting the heavy quark symmetry.
We focus on the $\bar{P}^{(*)}\Sigma_Q^{(*)}$ molecules with S-wave orbital angular momentum for simplicity.
The effect of tensor force by the S-D mixing is important for OPEP.
However, we do not include D-wave states complicating the system because
the S-wave channel is enough to see the spin decomposition to
the heavy quark spin and the other spin of the $\bar{P}^{(*)}\Sigma_Q^{(*)}$ molecules.
Our purpose in this paper is to demonstrate the HQS multiplet of $\bar{P}^{(*)}\Sigma_Q^{(*)}$.
Thus, we study the simple S-wave case in the present study.
Since the heavy quark spin and the other spin are separately
conserved by the heavy quark spin symmetry,
the heavy meson-baryon molecular basis is not suitable to discuss the
structure of HQS multiplet.
It is convenient to deal with the corresponding spin structure with appropriate basis.
Thus,
we define the light cloud spin (LCS) basis as a suitable basis to study the HQS multiplet structure and discuss that what types of HQS 
multiplets can exist under the OPEP.

This paper is organized as follows.
In Sec.\ref{sec:pot}, we construct the one-pion exchange potential in the hadronic molecular (HM) basis.
The basis transformation from the HM basis to the LCS basis is discussed in Sec.\ref{sec:HQS multiplet}.
We show the numerical result in Sec.\ref{sec:Numerical result}.
Finally, we summarize the work in this paper and discuss the result in Sec.\ref{sec:summary}.


\section{Potential}
\label{sec:pot}
In this section, we construct the OPEP for $\bar{P}^{(*)}\Sigma_Q^{(*)}$ molecular states based on the heavy quark symmetry and the chiral symmetry.
The $\bar{P}^{(*)}$ mesons and pion interaction Lagrangian is given by 
\cite{Falk:1991nq, Wise:1992hn, Cho:1992gg, Yan:1992gz, Falk:1992cx}
\begin{align}
	\mathcal{L}_{HH\pi} &= g\Tr\left[ \bar{H}H\gamma_{\mu}\gamma_5A^{\mu} \right]~.
\end{align}
The heavy meson doublet field $H$ is
\begin{align}
	H &= \frac{1+\vsla}{2}\left[ P_{\mu}^{*}\gamma^{\mu} + iP\gamma_{5} \right]\  \label{def H}.
\end{align}
$P$ and $P^*$ are pseudoscalar meson and vector meson fields in the HQS doublet.
The axial vector current for the pion is given by
\begin{align}
	A_{\mu} = \frac{i}{2}\left( \xi^{\dagger}\del_{\mu}\xi - \xi\del_{\mu}\xi^{\dagger} \right)~,
\end{align}
where $\xi = \exp(i\hat{\pi}/\sqrt{2}f_{\pi})$.
The pion decay constant is $f_{\pi}=92.4$ MeV and the pion field $\hat{\pi}$ is defined by
\begin{align}
	\hat{\pi} = \left(
	\begin{array}{cc}
		\pi^0/\sqrt{2} & \pi^+ \\
		\pi^- & -\pi^0/\sqrt{2}
	\end{array}
	\right)~.
\end{align}
The coupling constant $g$ is determined as $|g|=0.59$ from the decay of $D^* \to D\pi$\cite{Olive:2016xmw}.

The $\Sigma_Q^{(*)}$ baryons and pion interaction Lagrangian is given by \cite{Yan:1992gz, Liu:2011xc}
\begin{align}
	\mathcal{L}_{BB\pi} = \frac{3}{2}g_1iv_{\sigma}\epsilon^{\mu\nu\rho\sigma} \textrm{Tr} \left[ \bar{S}_{\mu}A_{\nu}S_{\rho} \right]~.
\end{align}
The superfield $S_{\mu}$ for $\Sigma_Q$ and $\Sigma_Q^*$ is represented as
\begin{align}
	S_{\mu} = \hat{\Sigma}_{Q\mu}^* - \sqrt{\frac{1}{3}}\left( \gamma_{\mu}+v_{\mu} \right)\gamma_5\hat{\Sigma}_Q~.
\end{align}
The heavy baryon fields $\hat{\Sigma}_{Q(\mu)}^{(*)}$ are defined by 
\begin{align}
	\hat{\Sigma}_{Q(\mu)}^{(*)} = \left(
	\begin{array}{cc}
		\Sigma_{Q(\mu)}^{(*)++} & \frac{1}{\sqrt{2}}\Sigma_{Q(\mu)}^{(*)+} \\
		\frac{1}{\sqrt{2}}\Sigma_{Q(\mu)}^{(*)+} & \Sigma_{Q(\mu)}^{(*)0}
	\end{array}
	\right)~.
\end{align}
$\Sigma_Q$ and $\Sigma_{Q\mu}^*$ are spin 1/2 and 3/2 baryon fields in the HQS doublet.
For the coupling constant $g_1$, we use $g_1=(\sqrt{8}/3)g_4$ and $g_4=0.999$ estimated in Ref. \cite{Liu:2011xc}.
The coupling $g_4$ is determined by the decay of $\Sigma_c^* \to \Lambda_c\pi$ and its sign follows the quark model estimation.

We construct the one pion exchange potential using the above Lagrangians.
At each vertex, we introduce a cutoff parameter $\Lambda$
via the monopole type form factor
\begin{align}
	F(q) = \frac{\Lambda^2 - m_{\pi}^2}{\Lambda^2 + |\vec{q}|^2}~,
\end{align}
where $m_{\pi}$ is a mass of the exchanging pion, and $\vec{q}$ is its momentum.
We use the same cutoff for $\bar{P}^{(*)}\bar{P}^{(*)}\pi$ and $\Sigma_Q^{(*)}\Sigma_Q^{(*)}\pi$ vertices for simplicity, and fix the value of cutoff $1000$ MeV and $1500$ MeV. 

In the present analysis, we concentrate on the 
S-wave $\bar{P}^{(*)}\Sigma_Q^{(*)}$ molecular states 
to clarify 
their HQS multiplet structures.
In the Hadronic Molecule (HM) basis,
the spin structures of molecular states are described by the product of
meson-baryon spins.
Then, the possible spins of the $\bar{P}^{(*)}\Sigma_Q^{(*)}$ states are
\begin{align}
    \bar{P}\Sigma_Q = 
    \left[\bar{Q}q\right]_0 \otimes \left[Q[d]_1\right]_{1/2} &= \frac{1}{2}~,\label{eq:spinPSigmaQ} \\
    \bar{P}\Sigma_Q^* = 
    \left[\bar{Q}q\right]_0 \otimes \left[Q[d]_1\right]_{3/2} &= \frac{3}{2}~,\label{eq:spinPSigmaQstar} \\
    \bar{P}^*\Sigma_Q = 
    \left[\bar{Q}q\right]_1 \otimes \left[Q[d]_1\right]_{1/2} &= \frac{1}{2} \oplus \frac{3}{2}~,\label{eq:spinPstarSigmaQ} \\
    \bar{P}^*\Sigma_Q^* = 
    \left[\bar{Q}q\right]_1 \otimes \left[Q[d]_1\right]_{3/2} &= \frac{1}{2} \oplus \frac{3}{2} \oplus \frac{5}{2}~,\label{eq:spinPstarSigmaQstar}
\end{align}
where
$Q$, $\bar{Q}$, $q$ and $d$ stand for a heavy quark, heavy antiquark,
light quark and diquark in $\Sigma_Q^{(*)}$ baryon, respectively, and
the index $j$ of $[\alpha]_j$ means the spin of $\alpha$.
The wavefunctions and OPEPs for each spin state are
\begin{align}
	\psi^{\textrm{HM}}_{1/2^-} &= \left(
	\begin{array}{l}
		\left| \bar{P}\Sigma_Q \right\rangle_{1/2^-} \\
		\left| \bar{P}^*\Sigma_Q \right\rangle_{1/2^-} \\
		\left| \bar{P}^*\Sigma_Q^* \right\rangle_{1/2^-}
	\end{array}
	\right)~, \\
	V^{\textrm{HM}}_{\pi, 1/2^-}(r) &= \frac{gg_1}{f_{\pi}^2}\left(
	\begin{array}{ccc}
		0 & -\frac{1}{\sqrt{3}} & \frac{1}{\sqrt{6}} \vspace{2pt} \\
		-\frac{1}{\sqrt{3}} & \frac{2}{3} & \frac{1}{3\sqrt{2}} \vspace{2pt} \\
		\frac{1}{\sqrt{6}} & \frac{1}{3\sqrt{2}} & \frac{5}{6}
	\end{array}
	\right)C_{\pi}(r)~,
\end{align}
\begin{align}
	\psi^{\textrm{HM}}_{3/2^-} &= \left(
	\begin{array}{l}
		\left| \bar{P}\Sigma_Q^* \right\rangle_{3/2^-} \\
		\left| \bar{P}^*\Sigma_Q \right\rangle_{3/2^-} \\
		\left| \bar{P}^*\Sigma_Q^* \right\rangle_{3/2^-}
	\end{array}
	\right)~, \\
	V^{\textrm{HM}}_{\pi, 3/2^-}(r) &= \frac{gg_1}{f_{\pi}^2}\left(
	\begin{array}{ccc}
		0 & -\frac{1}{2\sqrt{3}} & -\frac{5}{2\sqrt{15}} \vspace{2pt} \\
		-\frac{1}{2\sqrt{3}} & -\frac{1}{3} & \frac{5}{6\sqrt{5}} \vspace{2pt} \\
		-\frac{5}{2\sqrt{15}} & \frac{5}{6\sqrt{5}} & \frac{1}{3}
	\end{array}
	\right)C_{\pi}(r)~, 
\end{align}
\begin{align}
	\psi^{\textrm{HM}}_{5/2^-} &= \left(
	\begin{array}{l}
		\left| \bar{P}^*\Sigma_Q^* \right\rangle_{5/2^-}
	\end{array}
	\right)~, \\
	V^{\textrm{HM}}_{\pi, 5/2^-}(r) &= -\frac{gg_1}{2f_{\pi}^2}C_{\pi}(r)~.
\end{align}
The function $C_{\pi}(r)$ is defined as
\begin{align}
	C_{\pi}(r) = \frac{m_{\pi}^2}{4\pi}\left[ \frac{e^{-m_{\pi}r} - e^{-\Lambda r}}{r} - \frac{\Lambda^2 - m_{\pi}^2}{2\Lambda}e^{-\Lambda r} \right]~.
\end{align}
It should be noted that we 
subtract 
the contact terms from the potential.


\section{HQS multiplet structure of $\bar{P}^{(*)}\Sigma_Q^{(*)}$}
\label{sec:HQS multiplet}
We construct the OPEP for HM base in Sec.\ref{sec:pot}. 
However it is inconvenient to see the structure of HQS multiplet.
In this section,
we introduce the Light Cloud Spin (LCS) basis, where the spin
structure of $Q\bar{Q}qd$ states is divided into the heavy quark spin $[Q\bar{Q}]_S$
and light cloud spin $[q[d]_1]_j$.
It is a natural spin description in the heavy hadron systems,
because heavy quark spin and light cloud spin are separately conserved
in heavy quark effective theory.

Here, 
we treat the HQS structure of doubly heavy system in the following manner:  The
pentaquark 
as a bound state of $\bar{P}^{(*)}$ and $\Sigma_Q^{(*)}$ is labeled by 
the velocity $v$ of the pentaquark. 
It is natural 
to assume 
that both 
$\bar{P}^{(*)}$ and $\Sigma_Q^{(*)}$ have the same velocity $v$.

The spin structures of $\bar{P}^{(*)}\Sigma_Q^{(*)}$ molecular states in LCS basis are given by
\begin{align}
    \left[\bar{Q}Q\right]_0 \otimes \left[q[d]_1\right]_{1/2} &= \frac{1}{2}
     \hspace{5mm}({\rm singlet})~, \label{eq:spin structure 1/2 singlet} \\
    \left[\bar{Q}Q\right]_0 \otimes \left[q[d]_1\right]_{3/2} &= \frac{3}{2}
     \hspace{5mm}({\rm singlet})~, \label{eq:spin structure 3/2 singlet} \\
    \left[\bar{Q}Q\right]_1 \otimes \left[q[d]_1\right]_{1/2} &= \frac{1}{2} \oplus \frac{3}{2}
     \hspace{5mm}({\rm doublet})~, \label{eq:spin structure 1/2-3/2 doublet} \\
    \left[\bar{Q}Q\right]_1 \otimes \left[q[d]_1\right]_{3/2} &= \frac{1}{2} \oplus \frac{3}{2} \oplus \frac{5}{2}
      \hspace{5mm}({\rm triplet})~. \label{eq:spin structure 1/2-3/2-5/2 triplet}
\end{align}
There are four types of  HQS multiplets, spin $1/2$ singlet, spin $3/2$
singlet, spin $(1/2, 3/2)$ doublet and spin $(1/2, 3/2, 5/2)$ triplet
which are classified by the heavy quark spin $S=0,1$ and the light
cloud spin $j=1/2,3/2$.

Using unitary transformation matrices, we translate the basis from HM basis to LCS basis.
For spin $1/2$,
\begin{align}
	\psi^{\textrm{LCS}}_{1/2^-} &= U_{1/2^-}^{-1} \psi^{\textrm{HM}}_{1/2^-} \non \\
	 &= \left(
	\begin{array}{l}
		\left| \left[ \bar{Q}Q \right]_0 \otimes \left[ q[d]_1 \right]_{1/2} \right\rangle_{1/2^-}^{\textrm{singlet}} \\
		\left| \left[ \bar{Q}Q \right]_1 \otimes \left[ q[d]_1 \right]_{1/2} \right\rangle_{1/2^-}^{\textrm{doublet}} \\
		\left| \left[ \bar{Q}Q \right]_1 \otimes \left[ q[d]_1 \right]_{3/2} \right\rangle_{1/2^-}^{\textrm{triplet}}
	\end{array}
 \right)~,
 \label{eq:LCSwave1/2}
\end{align}
\begin{align}
	V^{\textrm{LCS}}_{\pi, 1/2^-}(r)  
	&= U_{1/2^-}^{-1} V_{1/2^-}^{\textrm{HM}} U_{1/2^-} \non \\
	&= \frac{gg_1}{f_{\pi}^2}\left(
	\begin{array}{ccc}
		1 & 0 & 0 \\
		0 & 1 & 0 \\
		0 & 0 & -\frac{1}{2}
	\end{array}
	\right)C_{\pi}(r)~.
	\label{eq:LCSpot1/2}
\end{align}
For spin $3/2$,
\begin{align}
	\psi^{\textrm{LCS}}_{3/2^-} &= U_{3/2^-}^{-1} \psi^{\textrm{HM}}_{3/2^-} \non \\
	&= \left(
	\begin{array}{l}
		\left| \left[ \bar{Q}Q \right]_0 \otimes \left[ q[d]_1 \right]_{3/2} \right\rangle_{3/2^-}^{\textrm{singlet}} \\
		\left| \left[ \bar{Q}Q \right]_1 \otimes \left[ q[d]_1 \right]_{1/2} \right\rangle_{3/2^-}^{\textrm{doublet}} \\
		\left| \left[ \bar{Q}Q \right]_1 \otimes \left[ q[d]_1 \right]_{3/2} \right\rangle_{3/2^-}^{\textrm{triplet}}
	\end{array}
	\right)~,
\end{align}
\begin{align}
V^{\textrm{LCS}}_{\pi, 3/2^-}(r) &= U_{3/2^-}^{-1} V_{3/2^-}^{\textrm{HM}} U_{3/2^-} \non \\
	&= \frac{gg_1}{f_{\pi}^2}\left(
	\begin{array}{ccc}
		-\frac{1}{2} & 0 & 0 \\
		0 & 1 & 0 \\
		0 & 0 & -\frac{1}{2}
	\end{array}
	\right)C_{\pi}(r)~.
	\label{eq:LCSpot3/2}
\end{align}
For spin $5/2$,
\begin{align}
\psi^{\textrm{LCS}}_{5/2^-} &= U_{5/2^-}^{-1} \psi^{\textrm{HM}}_{5/2^-} \non \\
	&= \left(
	\begin{array}{l}
		\left| \left[ \bar{Q}Q \right]_1 \otimes \left[ q[d]_1 \right]_{3/2} \right\rangle_{5/2^-}^{\textrm{triplet}}
	\end{array}
\right)~,
\end{align}
\begin{align}
	V^{\textrm{LCS}}_{\pi, 5/2^-}(r) &= U_{5/2^-}^{-1} V_{5/2^-}^{\textrm{HM}} U_{5/2^-} \non \\
	&= -\frac{gg_1}{2f_{\pi}^2}C_{\pi}(r)~.
	\label{eq:LCSpot5/2}
\end{align}
Here, we call the components labeled by (singlet, doublet, or
triplet) of $\psi^{\textrm{LCS}}_{J^P}$ as spin $J$ (singlet, doublet, or
triplet) state.
For instance, the first component of $\psi^{\textrm{LCS}}_{1/2^-}$ in
Eq.~\eqref{eq:LCSwave1/2}, i.e. $\left| \left[ \bar{Q}Q \right]_0
\otimes \left[ q[d]_1 \right]_{1/2}
\right\rangle_{1/2^-}^{\textrm{singlet}}$ is called spin 1/2 singlet state.
The transformation matrices determined by Clebsch-Gordan coefficient of spin reconstruction are given by
\begin{align}
U_{1/2^-} = \left(
	\begin{array}{ccc}
		\frac{1}{2} & -\frac{1}{2\sqrt{3}} & \frac{2}{\sqrt{6}} \\
		-\frac{1}{2\sqrt{3}} & \frac{5}{6} & \frac{2}{3\sqrt{2}} \\
		\frac{2}{\sqrt{6}} & \frac{2}{3\sqrt{2}} & -\frac{1}{3}
	\end{array}
\right)~,
\end{align}
\begin{align}
U_{3/2^-} = \left(
	\begin{array}{ccc}
		\frac{1}{2} & -\frac{1}{\sqrt{3}} & \frac{\sqrt{15}}{6} \\
		-\frac{1}{\sqrt{3}} & \frac{1}{3} & \frac{\sqrt{5}}{3} \\
		\frac{\sqrt{15}}{6} & \frac{\sqrt{5}}{3} & \frac{1}{6}
	\end{array}
\right)~,
\end{align}
\begin{align}
U_{5/2^-} = 1~.
\end{align}

The potential matrices in LCS basis are diagonalized corresponding to
the HQS multiplet components.
We find the particular values of the matrix elements of the OPEP; $+1$ for
$\left[ \bar{Q}Q \right]_0 \otimes \left[ q[d]_1 \right]_{1/2}$ and
$\left[ \bar{Q}Q \right]_1 \otimes \left[ q[d]_1 \right]_{1/2}$, and
$-2$ for $\left[ \bar{Q}Q \right]_0 \otimes \left[ q[d]_1 \right]_{3/2}$
and $\left[ \bar{Q}Q \right]_1 \otimes \left[ q[d]_1 \right]_{3/2}$.
Hence, these components play a different role, either an attraction or
a repulsion,
depending on the whole sign of the potential.


\section{Numerical result}
\label{sec:Numerical result}
Before solving coupled channel Schr{\"o}dinger equations under the LCS basis potential, 
let us discuss
the sign assignment of a coupling constant of the heavy meson-pion interaction, $|g|=0.59$~.
In the usual case, its sign is taken as plus following
quark models.
However, only the absolute value is determined by the decay of $D^* \to
D\pi$~\cite{Olive:2016xmw}, and
the sign of $g$ is not determined\footnote{It could be argued that the
relative sign of $g$ and $g_1$ is not determined.}.

This sign assignment is important in the present study.
For example, the coefficients of the HQS singlet and doublet component
are $+1$ in the spin $1/2$ potential of LCS basis in
Eq.~\eqref{eq:LCSpot1/2}.
Thus, these potentials play as a repulsive one
when we assign $g=+0.59$, but they are the attractive potentials when we choose $g=-0.59$.
On the other hand, the HQS triplet
with the coefficient $-1/2$ has the
attractive potential for $g=+0.59$ and repulsive potential for $g=-0.59$.
It is to say that the sign of the coupling constant
(the interaction models in general)
determines
which multiplets have 
the attractive potential.
We calculate the cases with both signs of $g$ to study the
behavior of the attractive multiplets in this section.

\subsection{Result in case of $g=+0.59$}
\begin{figure}[tbp]
\begin{center}
\includegraphics[bb = 0 0 640 384, width=0.45\textwidth]{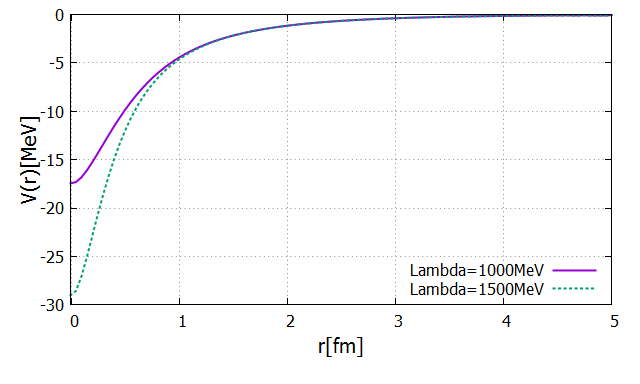}
\caption{Attractive potential $V(r) = -\frac{gg_1}{2\fpi^2}C_{\pi}$ for $\Lambda = 1000$ (purple solid curve) and $1500$ MeV (green dotted curve), where $g=0.59$, $g_1=0.942$ and $f_{\pi}=92.4$MeV.}
\label{fig:potential_plus}
\end{center}
\end{figure}

When we assign as $g=+0.59$, the HQS multiplets which have attractive potential are $J^P=3/2^-$ singlet and $J^P=(1/2^-, 3/2^-, 5/2^-)$ triplet.
The potential is written as
\begin{align}
	V(r) = -\frac{gg_1}{2f_{\pi}^2}C_{\pi}(r)~,
\end{align}
and we show it in Figure \ref{fig:potential_plus}.

Firstly, we show the results obtained by solving the Schr{\"o}dinger
equation with preserving the heavy quark spin symmetry.
We define the spin averaged mass for $\bar{P}^{(*)}$ mesons and $\Sigma_Q^{(*)}$ baryons as 
\begin{align}
	M_{\bar{P}{\textrm{ave}}} &= \frac{M_{\bar{P}} + 3M_{\bar{P}^*}}{4}~, \\
	M_{\Sigma_Q{\textrm{ave}}} &= \frac{2M_{\Sigma_Q} + 4M_{\Sigma_Q^*}}{6}~,
\end{align}
to deal with the degeneracy of the HQS doublet meson and baryon, respectively.
The masses of relevant 
charmed and bottomed hadrons are shown in Table \ref{tab:hadron masses}.
\begin{table}[btp]
\centering
\caption{Masses 
of relevant charmed and bottomed 
hadrons~\cite{Olive:2016xmw}.}
\begin{tabular}{c|cccc}\hline
	 & $\bar{D}$ & $\bar{D}^*$ & $B$ & $B^*$ \\
	Mass[MeV] & $1867.21$ & $2008.56$ & $5279.48$ & $5324.65$ \\ \hline 
	 & $\Sigma_c$ & $\Sigma_c^*$ & $\Sigma_b$ & $\Sigma_b^*$ \\
	Mass[MeV] & $2453.54$ & $2518.13$ & $5813.4$ & $5833.6$ \\ \hline 
\end{tabular}
\label{tab:hadron masses}
\end{table}
The spin averaged reduced mass is defined as 
\begin{align}
	\mu_{\textrm{ave}} = \frac{M_{\bar{P}{\textrm{ave}}} M_{\Sigma_Q{\textrm{ave}}}}{M_{\bar{P}{\textrm{ave}}} + M_{\Sigma_Q{\textrm{ave}}}}~.
\end{align}
When $\mu_{\textrm{ave}}=1.102, 1.474, 1.699$ and 2.779 GeV, the spin averaged masses of $\bar{D}^{(*)}\Sigma_c^{(*)}$, $\bar{D}^{(*)}\Sigma_b^{(*)}$, $B^{(*)}\Sigma_c^{(*)}$ and $B^{(*)}\Sigma_b^{(*)}$ are reproduced, respectively.
We solve the Schr{\"o}dinger equation with 
keeping 
the heavy quark spin symmetry
by changing the mass parameter $\mu_{\textrm{ave}}$ from 1 GeV to 100 GeV.
To obtain the bound state solutions, we use Gaussian expansion method \cite{Hiyama:2003cu}.

\begin{figure}[tbp]
\begin{center}
\includegraphics[bb = 0 0 640 384, width=0.45\textwidth]{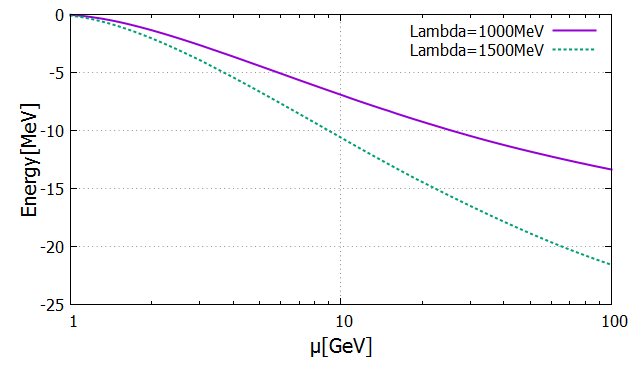}
\caption{Obtained binding energies for $\Lambda = 1000$ (purple solid curve) and $1500$ MeV  (green dotted curve) with $g=+0.59$.
The energy is measured from the threshold of $\bar{P}_{\rm ave}\Sigma_{Q {\rm ave}}$.
The mass parameter $\mu_{\rm ave}$ is changed from $1$GeV to $100$GeV.
}
\label{fig:BE_common_plus}
\end{center}
\end{figure}

The results of $\Lambda=1000$ and $1500$ MeV are shown in Figure \ref{fig:BE_common_plus}.
All four states, spin 3/2 singlet and spin (1/2, 3/2, 5/2) triplet, are degenerate because of the heavy quark spin symmetry and their bound state solutions are obtained for all range of $\mu_{\rm ave}$.
 
\begin{figure}[!tbp]
\begin{center}
\includegraphics[bb = 0 0 640 384, width=0.5\textwidth]{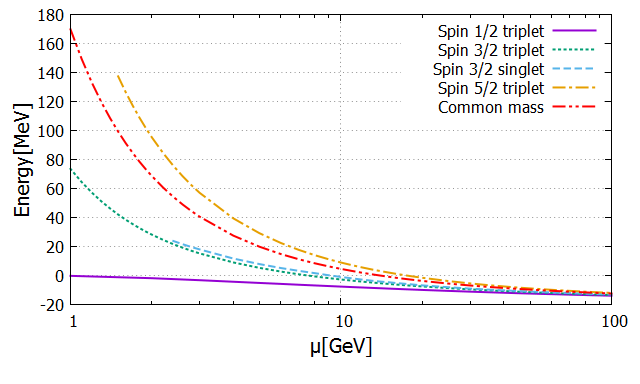}
\includegraphics[bb = 0 0 640 384, width=0.5\textwidth]{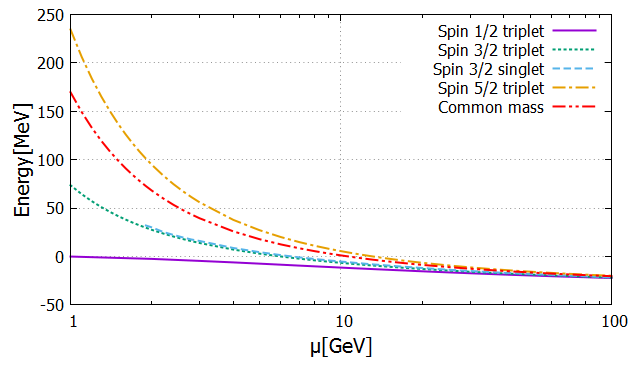}
\caption{ 
 Energies of the $\bar{P}^{(\ast)}\Sigma^{(\ast)}_Q$ states
with heavy quark spin symmetry breaking effect, obtained for
 $\Lambda=1000$ MeV (the upper figure) and $1500$ MeV (the lower figure)
 with $g=+0.59$.
These energies are measured from $\bar{P}\Sigma_Q$ threshold.
The purple solid, green dotted and yellow dashed-dotted 
curves 
are the energies of spin 
$(1/2, 3/2, 5/2)$ 
triplet states respectively and the light blue dashed curve is that of spin 3/2 singlet state.
For the sake of reference, 
we show the result for the case of keeping the heavy quark spin symmetry
 by the red dashed-dotted-dotted curve (Common mass).
}
\label{fig:Enmatome_plus}
\end{center}
\end{figure}

Next, we show the results including the effect of the heavy quark spin symmetry breaking.
The breaking is introduced by the nonzero mass difference
between the HQS multiplet, namely $\bar{P}$ and $\bar{P}^\ast$, and $\Sigma_{Q}$
and $\Sigma^\ast_{Q}$\footnote{The higher order terms of the effective
Lagrangians also break the heavy quark symmetry. However, we employ the
leading term of Lagrangians in this study.}.
To see the mass dependence of a binding energy, the heavy hadron masses
are parametrized as follows:

\begin{align}
	M_{\bar{P}} &= 2\mu + \frac{a}{2\mu} + \frac{w}{(2\mu)^2}~,
	\label{eq:mass fit 0meson} \\
	M_{\bar{P}^*} &= 2\mu + \frac{b}{2\mu} + \frac{x}{(2\mu)^2}~,
	\label{eq:mass fit 1meson} \\
	M_{\Sigma_Q} &= 2\mu + \frac{c}{2\mu} + \frac{y}{(2\mu)^2}~,
	\label{eq:mass fit 12baryon} \\
	M_{\Sigma_Q^*} &= 2\mu + \frac{d}{2\mu} + \frac{z}{(2\mu)^2}~,
	\label{eq:mass fit 32baryon}
\end{align}
where $\mu$ is a parameter corresponding to the reduced mass of $\bar{P}^{(*)}\Sigma_Q^{(*)}$ state.
The eight parameters of $a,b,c,d,w,x,y$ and $z$ are fixed to reproduce the eight hadron masses in Table \ref{tab:hadron masses}
by taking $\mu= 1.102$\,($2.779$)\,GeV for charm (bottom) sector. 
The values of these parameters are shown in Table \ref{tab:mass fit para}.
\begin{table*}[tbp]
\caption{ Values of 
parameters to include the 
effect of heavy quark spin symmetry breaking 
in Eqs.(\ref{eq:mass fit 0meson})-(\ref{eq:mass fit 32baryon}).}
\begin{center}
\begin{tabular}{cccccccc}\hline
	$a[\textrm{GeV}^2]$ & $b[\textrm{GeV}^2]$ & $c[\textrm{GeV}^2]$ & $d[\textrm{GeV}^2]$ &
	$w[\textrm{GeV}^3]$ & $x[\textrm{GeV}^3]$ & $y[\textrm{GeV}^3]$ & $z[\textrm{GeV}^3]$ \\ 
	-2.0798 & -1.8685 & 1.9889 & 2.0814 & 
	2.9468 & 3.1677 & -3.1729 & -3.0629 \\ \hline
\end{tabular}
\label{tab:mass fit para}
\end{center}
\end{table*}
Note that the 
charm (bottom) hadron masses are reproduced when we take $\mu=$1.102 (2.779) GeV, 
and the heavy quark spin symmetry is restored as the mass parameter $\mu$ increases.
The energies 
obtained by solving the Schr{\"o}dinger equations with the effect of
heavy quark spin symmetry breaking are shown in Figure
\ref{fig:Enmatome_plus}.
The labels in Figure~\ref{fig:Enmatome_plus}, e.g. Spin 1/2
triplet, are named as being it at the heavy quark limit.
For instance, the solid line named as Spin 1/2 triplet displays the
energy of the state which becomes the spin 1/2 triplet state at the
heavy quark limit.
We note that the components belonging to the same $J^P$ state can be mixed in the
finite hadron mass region as shown later, while they are not mixed at the heavy quark
limit.
The energies of the $\bar{D}^{(*)}\Sigma_c^{(*)}$,
$\bar{D}^{(*)}\Sigma_b^{(*)}$, $B^{(*)}\Sigma_c^{(*)}$ and
$B^{(*)}\Sigma_b^{(*)}$ states are correspond to the values at $\mu = 1.102$, $1.474$,
$1.699$ and $2.779$\,GeV,  respectively.

All 
four states are degenerate and the binding energy is 
$-13.7$ MeV for $\Lambda=1000$ MeV and $-22.3$ MeV for $\Lambda=1500$ MeV in heavy quark limit .
As $\mu$ becomes smaller, the degeneracy is solved.
At $\mu=1.102$ GeV, only two (three) states can be bound for $\Lambda=1000$ MeV ($1500$ MeV).

For the spin 1/2 and 3/2 states, each components is completely
separated in the heavy quark limit as shown in Eqs.~\eqref{eq:LCSpot1/2}
and \eqref{eq:LCSpot3/2}.
In the finite heavy hadron mass region, however, the kinetic term with
the nonzero mass splitting of the HQS multiplets gives
a mixing of the HQS singlet, doublet and triplet components.
The 
percentage of (singlet, doublet, triplet) components in wavefunctions 
for $\Lambda=1000$ MeV is shown in Table \ref{tab:ratio_plus}.
\begin{table*}[!htbp]
\caption{Percentage of (singlet, doublet, triplet) components in wavefunctions
of the spin $1/2$ and $3/2$ states in the case of $g=+0.59$ and $\Lambda=1000$ MeV.}
\begin{center}
\begin{tabular}{c|ccc}\hline
	$\mu[\textrm{GeV}]$ & Spin 1/2 triplet & Spin 3/2 triplet & Spin 3/2 singlet \\ \hline
	1 & $(0.8\%, 0\%, 99.2\%)$ &  $(1.6\%, 0\%, 98.4\%)$ & No bound state \\
	2 & $(0\%, 0\%, 100\%)$ &  $(0.9\%, 0\%, 99.1\%)$ & No bound state \\
	3 & $(0\%, 0\%, 100\%)$ &  $(0\%, 0\%, 100\%)$ & $(100\%, 0\%, 0\%)$ \\ \hline
\end{tabular}
\label{tab:ratio_plus}
\end{center}
\end{table*}
For $\mu \geq 3$ GeV, 
each component is 
perfectly separated.
These ratios are hardly changed even in the case of $\Lambda=1500$ MeV.
We can see that the effect of heavy quark spin symmetry breaking is small.

\subsection{Result in case of $g=-0.59$}
\begin{figure}[tbp]
\begin{center}
\includegraphics[bb = 0 0 640 384, width=0.45\textwidth]{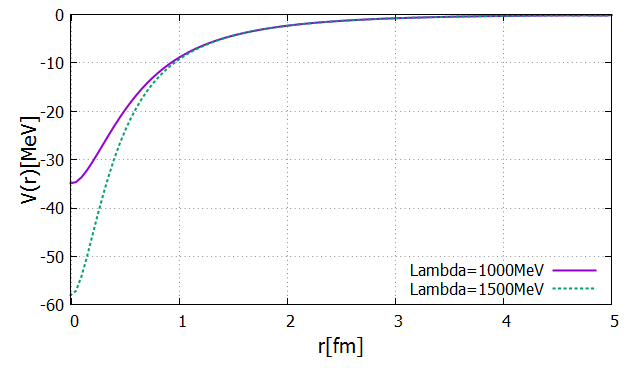}
\caption{Attractive 
potential $V(r) = \frac{gg_1}{\fpi^2}C_{\pi}$ for $\Lambda = 1000$ (purple solid curve) and $1500$ MeV (green dotted curve), where $g=-0.59$, $g_1=0.942$ and $f_{\pi}=92.4$MeV.}
\label{fig:potential_minus}
\end{center}
\end{figure}

In the case of $g=-0.59$~, the attractive multiplets are 
$J^P=1/2^-$ singlet and $J^P=(1/2^-, 3/2^-)$ doublet.
The potential is written as
\begin{align}
	V(r) = \frac{gg_1}{f_{\pi}^2}C_{\pi}(r)~,
\end{align}
and it is shown in Figure \ref{fig:potential_minus}.
This potential is twice deeper than that of $g=+0.59$~, 
and therefore we expect that the binding energy is larger.

\begin{figure}[tbp]
\begin{center}
\includegraphics[bb = 0 0 640 384, width=0.45\textwidth]{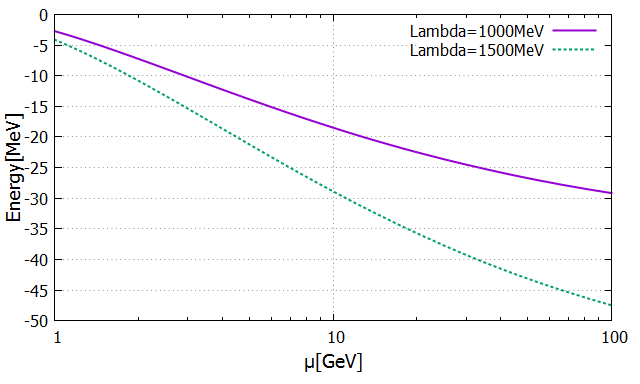}
\caption{Obtained binding energies for $\Lambda = 1000$ (purple solid curve)  and $1500$ MeV  (green dotted curve)  with $g=-0.59$.
The energy is measured from the threshold of $\bar{P}_{\rm ave}\Sigma_{Q {\rm ave}}$.
The mass parameter $\mu$ is changed from $1$GeV to $100$GeV.
}
\label{fig:BE_common_minus}
\end{center}
\end{figure}

\begin{figure}[!htbp]
\begin{center}
\includegraphics[bb = 0 0 640 384, width=0.5\textwidth]{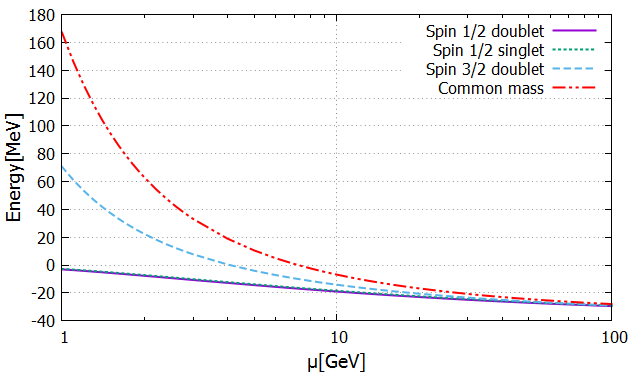}
\includegraphics[bb = 0 0 640 384, width=0.5\textwidth]{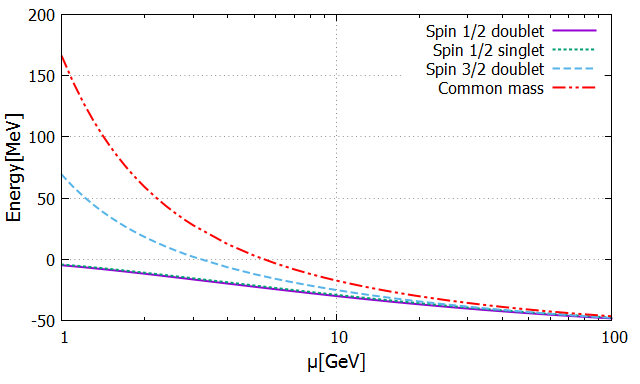}
\caption{
Energies obtained 
with heavy quark spin symmetry breaking for $\Lambda=1000$ MeV (the upper figure) and $1500$ MeV (the lower figure) with $g=-0.59$.
These energies are measured from $\bar{P}\Sigma_Q$ threshold.
The purple solid and light blue dashed curves are the energies of spin (1/2, 3/2) doublet states 
and the green dotted curve is that of spin 1/2 singlet state.
For the sake of reference, 
we show the result for the case of keeping the heavy quark spin symmetry by the red dashed-dotted-dotted curve.
}
\label{fig:Enmatome_minus}
\end{center}
\end{figure}

As in the case of $g=+0.59$~, 
we show the result that heavy quark spin symmetry is preserved in Figure \ref{fig:BE_common_minus} 
and the result that it is broken in Figure \ref{fig:Enmatome_minus}, 
for $g = - 0.59$~. 

Figure~\ref{fig:Enmatome_minus}
shows that all 
three states of spin $1/2$ singlet and $(1/2, 3/2)$ doublet are degenerate in the heavy quark limit and the binding energy is $-29.5$ MeV for $\Lambda=1000$ MeV and $-48.1$ MeV for $\Lambda=1500$ MeV,
which agree with the binding energies in the heavy quark limit shown in Figure~\ref{fig:BE_common_minus}. 
Unlike in the case of $g=+0.59$, all states are bound 
even at $\mu=1.102$ GeV corresponding to $\bar{D}^{(*)}\Sigma_c^{(*)}$
state and their binding energies are a few MeV.

The mixing ratio of wavefunction components for $\Lambda=1000$ MeV is shown in Table~\ref{tab:ratio_minus}.
\begin{table*}[!htbp]
\caption{Percentage of (singlet, doublet, triplet) components in wavefunctions
in the case of $g=-0.59$ and $\Lambda=1000$ MeV.}
\begin{center}
\begin{tabular}{c|ccc}\hline
	$\mu[\textrm{GeV}]$ & Spin 1/2 singlet & Spin 1/2 doublet & Spin 3/2 doublet \\ \hline
	1 & $(3.9\%, 96.1\%, 0\%)$ &  $(96.5\%, 3.4\%, 0.1\%)$ & $(0\%, 99.9\%, 0.1\%)$ \\
	2 & $(0.2\%, 99.8\%, 0\%)$ &  $(99.9\%, 0.1\%, 0\%)$ & $(0\%, 100\%, 0\%)$ \\
	3 & $(0\%, 100\%, 0\%)$ &  $(100\%, 0\%, 0\%)$ & $(0\%, 100\%, 0\%)$ \\ \hline
\end{tabular}
\label{tab:ratio_minus}
\end{center}
\end{table*}
The mixing ratio of the minor components induced by the heavy quark
symmetry breaking effect
is slightly larger than the case of $g=+0.59$, however it is still small.



\section{Summary and Discussions}
\label{sec:summary}

We 
showed that, in Sec.~\ref{sec:Numerical result}, 
the sign of a coupling constant of the heavy meson-pion interaction
determines which multiplets have the attractive potential.
In the case of $g=+0.59$~, spin $3/2$ HQS singlet and spin $(1/2, 3/2, 5/2)$ HQS triplet have the attractive potential.
On the other hand, in the case of $g=-0.59$~, spin $1/2$ HQS singlet and spin $(1/2, 3/2)$ HQS doublet have the attractive potential.

This classification is explained by the light cloud spin structure in 
Eqs.(\ref{eq:spin structure 1/2 singlet})-(\ref{eq:spin structure 1/2-3/2-5/2 triplet}).
The light cloud spin of spin $3/2$ singlet and spin $(1/2, 3/2, 5/2)$ triplet is $[q[d]_1]_{3/2}$ and 
that of spin $1/2$ singlet and spin $(1/2, 3/2)$ doublet is $[q[d]_1]_{1/2}$.
Because the pion exchange interaction 
\footnote{
We do not consider the tensor force in this study,
but it is also determined by the light cloud spin structure.
Not only the pion interaction, but also the other light meson interactions  
depend on the light cloud spin.
}
is coupled to the light quark spin, 
the difference of the attractive multiplet comes from the difference of the light cloud spin structure.
Moreover, we find the degeneracy of HQS singlet and triplet (singlet and
doublet) in the case of $g=+0.59$ ($-0.59$).
It is a natural result because
the OPEP does not depend on the heavy quark spin structure.

In the heavy quark limit, four (three) bound states exist for $g=+0.59$ ($-0.59$).
However, the heavy quark symmetry is broken for real  charm / bottom hadrons, 
so that all four (three) bound states may not exist in reality
as demonstrated in Sec.~\ref{sec:Numerical result}.  But we expect that 
there exist some HQS partners of $P_c$ like pentaquarks.
Especially, 
for bottom sector, 
the structure of HQS multiplet is more clearly
than for charm sector,
because the 
realization of the heavy quark symmetry is better. 
We expect the observation of the bottom pentaquarks to confirm the HQS multiplet structure of them.

The discussion in the LCS basis can be compared to the
quark model calculations, treating the constituent quarks as degrees of
freedom of the system.
In Refs.~\cite{Takeuchi:2016ejt,Yamaguchi:2017zmn}, the
short-range interaction in the $P_c$ pentaquarks are studied, which is derived
based on the quark cluster model.
The contributions from the color magnetic interaction of $c\bar{c}uud$ are evaluated, and
they find that the $c\bar{c}uud$ configurations having the other spin $j=3/2$
are important to
produce an attraction.
On the other hand, the configurations with $j=1/2$ give a repulsion.
In this study,
we also obtain that the states with $j=1/2$ and $j=3/2$ have 
a different role as shown in Eqs.~\eqref{eq:LCSpot1/2}, \eqref{eq:LCSpot3/2}
and  \eqref{eq:LCSpot5/2},
namely one is attractive and the other one is repulsive.
Thus, we find that a role of the interaction is characterized
by the light cloud spin in both of the quark model and the hadronic
molecular model.
It indicates that the discussion of the HQS multiplet structure can be
applied not only to the molecules, but also to the compact multiquark states.
In Ref.~\cite{Yamaguchi:2017zmn}, the $c\bar{c}uud$ potential for
$J^P=3/2^-$ with $(S,j)=(1,3/2)$ is stronger than that with
$(S,j)=(0,3/2)$.
This behavior also agrees with our results.

Focusing on the light cloud spin structure, there are constraints of the S-wave decay channel of the 
spin $3/2$ HQS singlet and spin $(1/2, 3/2, 5/2)$ HQS triplet.
Since their light cloud spin is given by $[q[d]_1]_{3/2}$, they cannot couple to the S-wave $[Q\bar{Q}]N$ and $\bar{P}^{(*)}\Lambda_Q$ states.
Here $[Q\bar{Q}]$, $N$ and $\Lambda_Q$ denote the heavy quarkonium, spin $1/2$ nucleon and HQS singlet heavy baryon like $\Lambda_c$, respectively.

Due to the heavy quark spin symmetry, heavy quark spin and light cloud spin are independently conserved.
Therefore, $[q[d]_1]_{3/2}$ having light cloud spin $3/2$ does not couple to the nucleon of spin $1/2$.
Moreover,
$[q[d]_1]_{3/2}$ cannot construct the diquark spin 0 by
the spin rearrangement.
So, $[q[d]_1]_{3/2}$ cannot couple to $\Lambda_Q$
with diquark spin 0 as well.
As a result, the S-wave decay channels to $[Q\bar{Q}]N$ and $\bar{P}^{(*)}\Lambda_Q$ 
from spin $3/2$ HQS singlet and spin $(1/2, 3/2, 5/2)$ HQS triplet 
are prohibited in the heavy quark limit.
There exist decay channels by the D-wave decay, however we expect that 
they are small. 

On the other hand, there are no constraint of S-wave decay
to $[Q\bar{Q}]N$ and $\bar{P}^{(*)}\Lambda_Q$
for spin
$1/2$ HQS singlet and spin $(1/2, 3/2)$ HQS doublet which have the light
cloud spin of $[q[d]_1]_{1/2}$ in the view of heavy quark spin symmetry.
These restrictions are independent of the model and derived only from heavy quark symmetry.
The difference in their S-wave decay channel restrictions should appear in the decay branching ratio of $Q\bar{Q}qqq$  pentaquark state.
We expect the measurement of the branching ratio to $[Q\bar{Q}]N$ and $\bar{P}^{(*)}\Lambda_Q$ to confirm the heavy quark symmetry in $P_c$-like pentaquarks.


\acknowledgments
The work of Y.S. is supported in part by JSPS Grant-in-Aid for JSPS Research Fellow No. JP17J06300. 
The work of M.H. is supported in part by 
JPSP KAKENHI
Grant Number 16K05345. 
The work of Y.Y. is supported in part by the Special Postdoctoral Researcher (SPDR) and iTHEMS Programs of RIKEN.

We are grateful to Sachiko Takeuchi, Atsushi Hosaka and Tetsuo Hyodo for useful comments and discussions.


\begin{thebibliography}{99}

\bibitem{Aaij:2015tga} 
  R.~Aaij {\it et al.} [LHCb Collaboration],
  Observation of $J/\psi p$ Resonances Consistent with Pentaquark States in $\Lambda_b^0 \to J/\psi K^- p$ Decays,
  Phys.\ Rev.\ Lett.\  {\bf 115}, 072001 (2015).

\bibitem{Aaij:2016phn} 
  R.~Aaij {\it et al.} [LHCb Collaboration],
  Model-independent evidence for $J/\psi p$ contributions to $\Lambda_b^0\to J/\psi p K^-$ decays,
  Phys.\ Rev.\ Lett.\  {\bf 117}, no. 8, 082002 (2016).

\bibitem{Aaij:2016ymb} 
  R.~Aaij {\it et al.} [LHCb Collaboration],
  Evidence for exotic hadron contributions to $\Lambda_b^0 \to J/\psi p \pi^-$ decays,
  Phys.\ Rev.\ Lett.\  {\bf 117}, no. 8, 082003 (2016).


\bibitem{Wu:2010jy} 
  J.~J.~Wu, R.~Molina, E.~Oset and B.~S.~Zou,
  Prediction of narrow $N^*$ and $\Lambda^*$ resonances with hidden charm above 4 GeV,
  Phys.\ Rev.\ Lett.\  {\bf 105}, 232001 (2010).

\bibitem{Yang:2011wz} 
  Z.~C.~Yang, Z.~F.~Sun, J.~He, X.~Liu and S.~L.~Zhu,
  The possible hidden-charm molecular baryons composed of anti-charmed meson and charmed baryon,
  Chin.\ Phys.\ C {\bf 36}, 6 (2012).

\bibitem{Wang:2011rga} 
  W.~L.~Wang, F.~Huang, Z.~Y.~Zhang and B.~S.~Zou,
  $\Sigma_c \bar{D}$ and $\Lambda_c \bar{D}$ states in a chiral quark model,
  Phys.\ Rev.\ C {\bf 84}, 015203 (2011)

\bibitem{Wu:2012md} 
  J.~J.~Wu, T.-S.~H.~Lee and B.~S.~Zou,
  Nucleon Resonances with Hidden Charm in Coupled-Channel Models,
  Phys.\ Rev.\ C {\bf 85}, 044002 (2012).


\bibitem{Chen:2015loa} 
  R.~Chen, X.~Liu, X.~Q.~Li and S.~L.~Zhu,
  Identifying exotic hidden-charm pentaquarks,
  Phys.\ Rev.\ Lett.\  {\bf 115}, no. 13, 132002 (2015).

\bibitem{He:2015cea} 
  J.~He,
  $\bar{D}\Sigma^*_c$ and $\bar{D}^*\Sigma_c$ interactions and the LHCb hidden-charmed pentaquarks,
  Phys.\ Lett.\ B {\bf 753}, 547 (2016).

\bibitem{Chen:2015moa} 
  H.~X.~Chen, W.~Chen, X.~Liu, T.~G.~Steele and S.~L.~Zhu,
  Towards exotic hidden-charm pentaquarks in QCD,
  Phys.\ Rev.\ Lett.\  {\bf 115}, no. 17, 172001 (2015).

\bibitem{Huang:2015uda} 
  H.~Huang, C.~Deng, J.~Ping and F.~Wang,
  Possible pentaquarks with heavy quarks,
  Eur.\ Phys.\ J.\ C {\bf 76}, no. 11, 624 (2016).

\bibitem{Roca:2015dva} 
  L.~Roca, J.~Nieves and E.~Oset,
  LHCb pentaquark as a $\bar{D}^*\Sigma_c-\bar{D}^*\Sigma_c^*$ molecular state,
  Phys.\ Rev.\ D {\bf 92}, no. 9, 094003 (2015).

\bibitem{Meissner:2015mza} 
  U.~G.~Mei{\ss}ner and J.~A.~Oller,
  Testing the $\chi_{c1}\, p$ composite nature of the $P_c(4450)$,
  Phys.\ Lett.\ B {\bf 751}, 59 (2015).

\bibitem{Xiao:2015fia} 
  C.~W.~Xiao and U.-G.~Mei{\ss}ner,
  $J/\psi(\eta_c)N$ and $\Upsilon(\eta_b)N$ cross sections,
  Phys.\ Rev.\ D {\bf 92}, no. 11, 114002 (2015).

\bibitem{Burns:2015dwa} 
  T.~J.~Burns,
  Phenomenology of $P_{c}$(4380)$^{+}$, $P_{c}$(4450)$^{+}$ and related states,
  Eur.\ Phys.\ J.\ A {\bf 51}, no. 11, 152 (2015).

\bibitem{Kahana:2015tkb} 
  D.~E.~Kahana and S.~H.~Kahana,
  LHCb $P_c^+$ Resonances as Molecular States,
  arXiv:1512.01902 [hep-ph].

\bibitem{Chen:2016heh} 
  R.~Chen, X.~Liu and S.~L.~Zhu,
  Hidden-charm molecular pentaquarks and their charm-strange partners,
  Nucl.\ Phys.\ A {\bf 954}, 406 (2016).

\bibitem{Chen:2016otp} 
  H.~X.~Chen, E.~L.~Cui, W.~Chen, X.~Liu, T.~G.~Steele and S.~L.~Zhu,
  QCD sum rule study of hidden-charm pentaquarks,
  Eur.\ Phys.\ J.\ C {\bf 76}, no. 10, 572 (2016).

\bibitem{Shimizu:2016rrd} 
  Y.~Shimizu, D.~Suenaga and M.~Harada,
  Coupled channel analysis of molecule picture of $P_{c}(4380)$,
  Phys.\ Rev.\ D {\bf 93}, no. 11, 114003 (2016).

\bibitem{Yamaguchi:2016ote} 
  Y.~Yamaguchi and E.~Santopinto,
  Hidden-charm pentaquarks as a meson-baryon molecule with coupled channels for $\bar{D}^{(\ast)}\Lambda_{\rm c}$ and $\bar{D}^{(\ast)}\Sigma^{(\ast)}_{\rm c}$,
  Phys.\ Rev.\ D {\bf 96}, no. 1, 014018 (2017).

\bibitem{He:2016pfa} 
  J.~He,
  Understanding spin parity of $P_c(4450)$ and $Y(4274)$ in a hadronic molecular state picture,
  Phys.\ Rev.\ D {\bf 95}, no. 7, 074004 (2017).

\bibitem{Ortega:2016syt} 
  P.~G.~Ortega, D.~R.~Entem and F.~Fern\'andez,
  LHCb pentaquarks in constituent quark models,
  Phys.\ Lett.\ B {\bf 764}, 207 (2017)

\bibitem{Azizi:2016dhy} 
  K.~Azizi, Y.~Sarac and H.~Sundu,
  Analysis of $P_c^+(4380)$ and $P_c^+(4450)$ as pentaquark states in the molecular picture with QCD sum rules,
  Phys.\ Rev.\ D {\bf 95}, no. 9, 094016 (2017).

\bibitem{Geng:2017hxc} 
  L.~Geng, J.~Lu and M.~P.~Valderrama,
  Scale Invariance in Heavy Hadron Molecules,
  arXiv:1704.06123 [hep-ph].




\bibitem{Maiani:2015vwa} 
  L.~Maiani, A.~D.~Polosa and V.~Riquer,
  The New Pentaquarks in the Diquark Model,
  Phys.\ Lett.\ B {\bf 749}, 289 (2015).

\bibitem{Lebed:2015tna} 
  R.~F.~Lebed,
  The Pentaquark Candidates in the Dynamical Diquark Picture,
  Phys.\ Lett.\ B {\bf 749}, 454 (2015).

\bibitem{Anisovich:2015cia} 
  V.~V.~Anisovich, M.~A.~Matveev, J.~Nyiri, A.~V.~Sarantsev and A.~N.~Semenova,
  Pentaquarks and resonances in the $pJ/\psi$ spectrum,
  arXiv:1507.07652 [hep-ph].

\bibitem{Li:2015gta} 
  G.~N.~Li, X.~G.~He and M.~He,
  Some Predictions of Diquark Model for Hidden Charm Pentaquark Discovered at the LHCb,
  JHEP {\bf 1512}, 128 (2015).

\bibitem{Wang:2015epa} 
  Z.~G.~Wang,
  Analysis of $P_c(4380)$ and $P_c(4450)$ as pentaquark states in the diquark model with QCD sum rules,
  Eur.\ Phys.\ J.\ C {\bf 76}, no. 2, 70 (2016).

\bibitem{Zhu:2015bba} 
  R.~Zhu and C.~F.~Qiao,
  Pentaquark states in a diquark-triquark model,
  Phys.\ Lett.\ B {\bf 756}, 259 (2016)

\bibitem{Santopinto:2016pkp} 
  E.~Santopinto and A.~Giachino,
  Compact pentaquark structures,
  Phys.\ Rev.\ D {\bf 96}, no. 1, 014014 (2017).

\bibitem{Wu:2017weo} 
  J.~Wu, Y.~R.~Liu, K.~Chen, X.~Liu and S.~L.~Zhu,
  Hidden-charm pentaquarks and their hidden-bottom and $B_c$-like partner states,
  Phys.\ Rev.\ D {\bf 95}, no. 3, 034002 (2017).

\bibitem{Hiyama:2018ukv} 
  E.~Hiyama, A.~Hosaka, M.~Oka and J.~M.~Richard,
  Quark model estimate of hidden-charm pentaquark resonances,
  arXiv:1803.11369 [nucl-th].


\bibitem{Takeuchi:2016ejt} 
  S.~Takeuchi and M.~Takizawa,
  The hidden charm pentaquarks are the hidden color-octet $uud$ baryons?,
  Phys.\ Lett.\ B {\bf 764}, 254 (2017).


\bibitem{Kubarovsky:2015aaa} 
  V.~Kubarovsky and M.~B.~Voloshin,
  Formation of hidden-charm pentaquarks in photon-nucleon collisions,
  Phys.\ Rev.\ D {\bf 92}, no. 3, 031502 (2015).


\bibitem{Eides:2017xnt} 
  M.~I.~Eides, V.~Y.~Petrov and M.~V.~Polyakov,
  Pentaquarks with hidden charm as hadroquarkonia,
  Eur.\ Phys.\ J.\ C {\bf 78}, no. 1, 36 (2018).

\bibitem{Scoccola:2015nia} 
  N.~N.~Scoccola, D.~O.~Riska and M.~Rho,
  Pentaquark candidates P$_c^+$(4380) and P$_c^+$(4450) within the soliton picture of baryons,
  Phys.\ Rev.\ D {\bf 92}, no. 5, 051501 (2015).


\bibitem{Yamaguchi:2017zmn} 
  Y.~Yamaguchi, A.~Giachino, A.~Hosaka, E.~Santopinto, S.~Takeuchi and M.~Takizawa,
  Hidden-charm and bottom meson-baryon molecules coupled with five-quark states,
  Phys.\ Rev.\ D {\bf 96}, no. 11, 114031 (2017).


\bibitem{Guo:2015umn} 
  F.~K.~Guo, U.~G.~Mei{\ss}ner, W.~Wang and Z.~Yang,
  How to reveal the exotic nature of the P$_c$(4450),
  Phys.\ Rev.\ D {\bf 92}, no. 7, 071502 (2015).
  
\bibitem{Liu:2015fea} 
  X.~H.~Liu, Q.~Wang and Q.~Zhao,
  Understanding the newly observed heavy pentaquark candidates,
  Phys.\ Lett.\ B {\bf 757}, 231 (2016).

\bibitem{Mikhasenko:2015vca} 
  M.~Mikhasenko,
  A triangle singularity and the LHCb pentaquarks,
  arXiv:1507.06552 [hep-ph].

\bibitem{Guo:2016bkl} 
  F.~K.~Guo, U.~G.~Mei{\ss}ner, J.~Nieves and Z.~Yang,
  Remarks on the $P_c$ structures and triangle singularities,
  Eur.\ Phys.\ J.\ A {\bf 52}, no. 10, 318 (2016).

\bibitem{Bayar:2016ftu} 
  M.~Bayar, F.~Aceti, F.~K.~Guo and E.~Oset,
  A Discussion on Triangle Singularities in the $\Lambda_b \to J/\psi K^{-} p$ Reaction,
  Phys.\ Rev.\ D {\bf 94}, no. 7, 074039 (2016).


\bibitem{Takeuchi:2014rsa} 
  S.~Takeuchi, K.~Shimizu and M.~Takizawa,
  On the origin of the narrow peak and the isospin symmetry breaking of the $X$(3872),
  PTEP {\bf 2014}, no. 12, 123D01 (2014),
  Erratum: [PTEP {\bf 2015}, no. 7, 079203 (2015)].

\bibitem{Isgur:1989vq}
 N.~Isgur and M.~B.~Wise,
 Weak Decays of Heavy Mesons in the Static Quark Approximation,
 Phys.\ Lett.\ B {\bf 232} (1989) 113. 
	
\bibitem{Isgur:1989ed}
 N.~Isgur and M.~B.~Wise,
 Weak Transition Form-factors Between Heavy Mesons,
 Phys.\ Lett.\ B {\bf 237}, 527 (1990). 

\bibitem{Isgur:1991wq}
 N.~Isgur and M.~B.~Wise,
 Spectroscopy with heavy quark symmetry,
 Phys.\ Rev.\ Lett.\ {\bf 66} (1991) 1130. 

\bibitem{Neubert:1993mb} 
 M.~Neubert,
 Heavy quark symmetry,
 Phys.\ Rept.\ {\bf 245}, 259 (1994).

\bibitem{Manohar:2000dt}
 A.~V.~Manohar and M.~B.~Wise,
 Heavy Quark Physics,
 (Cambridge University Press, Cambridge, 2000)
 
	
\bibitem{Yasui:2013vca} 
  S.~Yasui, K.~Sudoh, Y.~Yamaguchi, S.~Ohkoda, A.~Hosaka and T.~Hyodo,
  Spin degeneracy in multi-hadron systems with a heavy quark,
  Phys.\ Lett.\ B {\bf 727}, 185 (2013).

\bibitem{Yamaguchi:2014era} 
  Y.~Yamaguchi, S.~Ohkoda, A.~Hosaka, T.~Hyodo and S.~Yasui,
  Heavy quark symmetry in multihadron systems,
  Phys.\ Rev.\ D {\bf 91}, 034034 (2015).

\bibitem{Hosaka:2016ypm} 
 A.~Hosaka, T.~Hyodo, K.~Sudoh, Y.~Yamaguchi and S.~Yasui,
 Heavy Hadrons in Nuclear Matter,
 Prog.\ Part.\ Nucl.\ Phys.\ {\bf 96}, 88 (2017).
	

\bibitem{Falk:1991nq} 
  A.~F.~Falk,
  Hadrons of arbitrary spin in the heavy quark effective theory,
  Nucl.\ Phys.\ B {\bf 378}, 79 (1992).

\bibitem{Wise:1992hn} 
  M.~B.~Wise,
  Chiral perturbation theory for hadrons containing a heavy quark,
  Phys.\ Rev.\ D {\bf 45}, no. 7, R2188 (1992).

\bibitem{Cho:1992gg} 
  P.~L.~Cho,
  Chiral perturbation theory for hadrons containing a heavy quark: The Sequel,
  Phys.\ Lett.\ B {\bf 285}, 145 (1992).

\bibitem{Yan:1992gz} 
  T.~M.~Yan, H.~Y.~Cheng, C.~Y.~Cheung, G.~L.~Lin, Y.~C.~Lin and H.~L.~Yu,
  Heavy quark symmetry and chiral dynamics,
  Phys.\ Rev.\ D {\bf 46}, 1148 (1992)
  Erratum: [Phys.\ Rev.\ D {\bf 55}, 5851 (1997)].

\bibitem{Falk:1992cx} 
  A.~F.~Falk and M.~E.~Luke,
  Strong decays of excited heavy mesons in chiral perturbation theory,
  Phys.\ Lett.\ B {\bf 292}, 119 (1992).

\bibitem{Liu:2011xc} 
  Y.~R.~Liu and M.~Oka,
  $\Lambda_c N$ bound states revisited,
  Phys.\ Rev.\ D {\bf 85}, 014015 (2012).


\bibitem{Olive:2016xmw} 
  C.~Patrignani {\it et al.} [Particle Data Group],
  Review of Particle Physics,
  Chin.\ Phys.\ C {\bf 40}, no. 10, 100001 (2016).

\bibitem{Hiyama:2003cu} 
  E.~Hiyama, Y.~Kino and M.~Kamimura,
  Gaussian expansion method for few-body systems,
  Prog.\ Part.\ Nucl.\ Phys.\  {\bf 51}, 223 (2003).











\end{thebibliography}
\end{document}